\documentclass[aps,prl,amsmath, reprint,superscriptaddress]{revtex4-2}

\usepackage{graphicx}
\usepackage{bm}

\newcommand{\state}[1]{\left|#1\right\rangle}

\newcommand{\abs}[1]{\left|#1\right|}
\newcommand{\avg}[1]{\left\langle#1 \right\rangle}
\newcommand{\sUnit}{\mathrm{nV} \mathrm{cm}^{-1} \mathrm{Hz}^{-1/2}}

\begin{document}


\title{High-Sensitive Microwave Electrometry with Enhanced Instantaneous Bandwidth}


\author{Bowen Yang}
\affiliation{Key Laboratory of Quantum Optics and Center of Cold Atom Physics, Shanghai Institute of Optics and Fine Mechanics, Chinese Academy of Sciences, Shanghai 201800, China}
\affiliation{Center of Materials Science and Optoelectronics Engineering, University of Chinese Academy of Sciences, Beijing 100049, China}

\author{Yuhan Yan}
\affiliation{Key Laboratory of Quantum Optics and Center of Cold Atom Physics, Shanghai Institute of Optics and Fine Mechanics, Chinese Academy of Sciences, Shanghai 201800, China}
\affiliation{Department of Physics and Electronic Science, East China Normal University, Shanghai 200062, China}

\author{Xuejie Li}
\affiliation{Key Laboratory of Quantum Optics and Center of Cold Atom Physics, Shanghai Institute of Optics and Fine Mechanics, Chinese Academy of Sciences, Shanghai 201800, China}
\affiliation{Department of Physics and Electronic Science, East China Normal University, Shanghai 200062, China}

\author{Ling Xiao}
\affiliation{Key Laboratory of Quantum Optics and Center of Cold Atom Physics, Shanghai Institute of Optics and Fine Mechanics, Chinese Academy of Sciences, Shanghai 201800, China}

\author{Xiaolin Li}
\affiliation{School of Physics, East China University of Science and Technology}

\author{L. Q. Chen}
\email[]{lqchen@phy.ecnu.edu.cn}
\affiliation{Department of Physics and Electronic Science, East China Normal University, Shanghai 200062, China}
\affiliation{Shanghai Branch, Hefei National Laboratory, Shanghai 201315, China}

\author{Jianliao Deng}
\email[]{jldeng@siom.ac.cn}
\affiliation{Key Laboratory of Quantum Optics and Center of Cold Atom Physics, Shanghai Institute of Optics and Fine Mechanics, Chinese Academy of Sciences, Shanghai 201800, China}

\author{Huadong Cheng}
\email[]{chenghd@siom.ac.cn}
\affiliation{Key Laboratory of Quantum Optics and Center of Cold Atom Physics, Shanghai Institute of Optics and Fine Mechanics, Chinese Academy of Sciences, Shanghai 201800, China}
\affiliation{Center of Materials Science and Optoelectronics Engineering, University of Chinese Academy of Sciences, Beijing 100049, China}



\date{\today}

\begin{abstract}
Rydberg microwave (MW) sensors are superior to conventional antenna-based techniques because of their wide operating frequency range and outstanding potential sensitivity. Here, we demonstrate a Rydberg microwave receiver with a high sensitivity of $62\,\sUnit$ and broad instantaneous bandwidth of up to $10.2\,\mathrm{MHz}$. Such excellent performance was achieved by the amplification of one generated sideband wave induced by the strong coupling field in the six-wave mixing process of the Rydberg superheterodyne receiver, which was well predicted by our theory. Our system, which possesses a uniquely enhanced instantaneous bandwidth and high-sensitivity features that can be improved further, will promote the application of Rydberg microwave electrometry in radar and communication.
	
\end{abstract}


\maketitle


	Atomic systems, which are excellent quantum platforms, have achieved tremendous developments in precision quantum metrology such as magnetometry \cite{b_mag,b_mag2,b_mag3} and atomic frequency standards \cite{b_clk1, b_clk2, b_clk3}. Quantum sensing of MW electric fields has also gained considerable attention because of its potential applications in radar \cite{b_radar} and modern communication \cite{b_mc}. Rydberg atoms, which possess large dipole moments of MW transitions \cite{b_Rydberg}, are a major platform for providing highly sensitive microwave measurements\cite{b_sensor12_Np,b_sensor16,b_sensor17,b_sensor17_2,b_sensor20,b_sensor20_2,b_sensor20_Np,b_sensor21,b_sensor21_2,b_sensor21_3,b_sensor22}. The amplitude of the MW electric field has been detected using Autler-Townes (AT) splitting or ac Stark shifts induced by MW in the electromagnetically induced transparency (EIT) spectrum, providing a direct	International System of Units traceable measurement of MW \cite{b_sensor21,b_sensor22, b_at}. In addition, the ability to measure the	polarization \cite{b_vec,b_vec2}, phase \cite{b_sensor20_Np, b_phase}, and angle of arrival \cite{b_angle} of the MW field has been demonstrated.
	
	High sensitivity and broad instantaneous bandwidth are two core parameters for high-performance practical MW electrometry. To achieve high	sensitivity, a method called a superheterodyne receiver utilizing an extra	local dressed MW has been demonstrated, achieving unprecedented highly sensitive MW measurements with a sensitivity of up to $55\,\sUnit$ \cite{b_sensor20_Np}; however, the instantaneous bandwidth is limited to a few hundred kilohertz. For the instantaneous bandwidth, although a response time of $10\,\mathrm{ns}$ for MW pulses was demonstrated in \cite{b_time}, the bandwidth observed in most experiments on Rydberg communication receivers is below several megahertz \cite{b_band1,b_band2,b_band3,b_band4,b_band5}. The limitation of the transient atomic response to MW pulses results from dephasing mechanisms \cite{respone}. Recently, a large modulation bandwidth for MW communication in a Rydberg superheterodyne receiver was demonstrated with the discovery of an intrinsic six-wave mixing process \cite{arXiv_sixwaveSH}.	However, there has been no reported Rydberg MW sensor that can satisfy the high sensitivity and broad bandwidth requirements for practical applications, such as radar.
	
	In this letter, we demonstrate a Rydberg superheterodyne receiver with a sensitivity of $62\,\sUnit$ and an instantaneous bandwidth of up to $10.2\,\mathrm{MHz}$ based on a new	physical mechanism of enhanced atomic transient response, which was well predicted by our theoretical model. The physical mechanism for the enhancement of the atomic transient response is the increased emission efficiency of one generated sideband wave in the six-wave mixing process, which leads to a sensitive response of Rydberg systems to microwaves with a broad instantaneous bandwidth.

\begin{figure}
		\includegraphics[width=8.6cm]{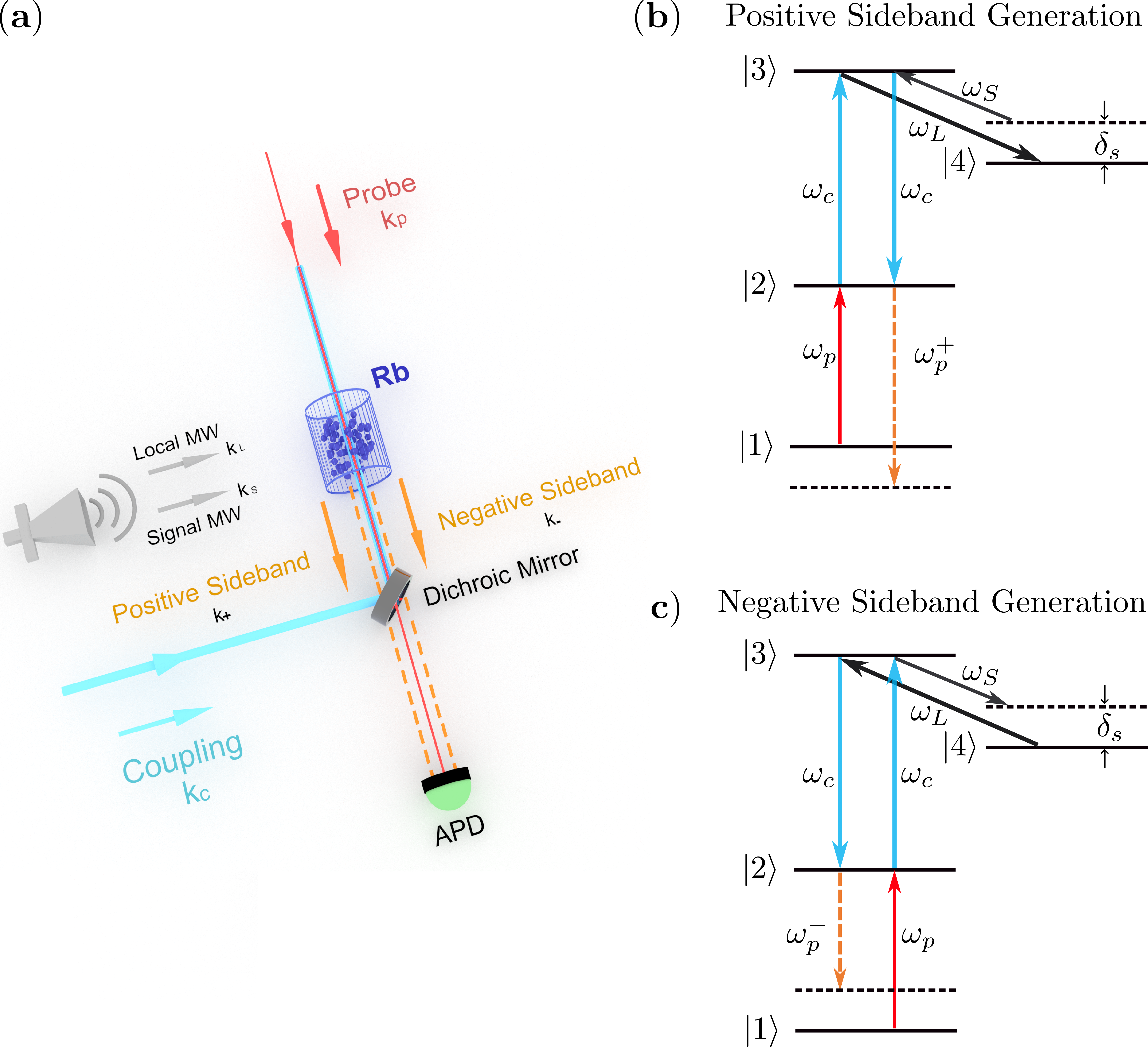}
		\caption{\label{expFig}Principle of Rydberg MW sensor. (a) Schematic of the experiment setup. The two laser beams counterpropagate through the vapor cell. The signal and local MW fields are emitted from the MW horn antenna. All fields are $\pi$ polarized and aligned in the same direction. As the probe light passes through rubidium vapor cell, in the presence of signal MW field, two sideband waves at $\omega_p\pm\delta_s$ are generated and mixed with the probe light, which is then recorded by the avalanche photodiode (APD), constructing a heterodyne detection. Panels (b) and (c) correspond to the generation of positive and negative sidebands in the two six-wave mixing processes, respectively. The levels $\state{1}$, $\state{2}$, $\state{3}$, and $\state{4}$ correspond to $^{87}\mathrm{Rb}$ ground state $5 S_{1/2}$, excited state $5 P_{3/2}$, Rydberg states $51D_{5/2}$, and $52P_{3/2}$, respectively. The frequencies of probe laser, coupling laser, and local and signal MWs are denoted as $\omega_p$, $\omega_c$, and $\omega_L$ and $\omega_S$, respectively. Two sideband waves at $\omega_p^+ = \omega_p+\delta_s$ and $\omega_p^-=\omega_p-\delta_s$ were generated by the two six-wave mixing processes, where $\delta_s = \omega_L -\omega_S$ is the frequency separation of the two MWs. The enhanced atomic transient response was achieved through the amplification of negative sideband wave in our system.}
	\end{figure}

	The experiment was performed on a $5$-cm long cylindrical room-temperature rubidium vapor cell, as shown in Fig.~\ref{expFig}(a). The beam for the probe field was derived from a 780 nm laser stabilized using modulation transfer spectroscopy, whereas the coupling field came from a 480 nm laser locked to a high-finesse Fabry-Perot cavity. The Gaussian $1/e^{2}$ beam radii of the probe and coupling light were $78.66\,\mathrm{\mu m}$ and $100.24\,\mathrm{\mu m}$, respectively. The power of the coupling laser was varied in the experiment, whereas that of the probe laser was maintained at $404\,\mathrm{nW}$, which corresponded to a peak Rabi frequency of $\Omega _{p}/2\pi=5.53\,\mathrm{MHz}$,	because a lower probe Rabi frequency leads to a faster	instantaneous response \cite{respone}. The generated optical signals and transmitted probe laser were recorded using an avalanche photodiode (APD) with a bandwidth of $10\,\mathrm{MHz}$ (Thorlasbs APD410A) and analyzed using a spectrum analyzer (Keysight N9020A MXA) to obtain signal amplitudes via a superheterodyne. The signal and local MWs were generated using two analog signal generators (Agilent E8257D), combined using a power combiner, and transmitted through a single horn antenna to free space. The local MW resonantly coupled Rydberg states at a frequency of $16.0304\,\mathrm{GHz}$. The reference clocks of the two MW sources and spectrum analyzer were synchronized with an active hydrogen clock to maintain frequency stability. The details of the Rabi frequency calibrations of the two MWs and lasers are presented in \cite{sup}.
	
	The energy levels for the six-wave mixing processes in the Rydberg superheterodyne receiver are shown in Figs.~\ref{expFig}(b) and (c). The probe ($\omega _{p}$), coupling ($\omega _{c}$), and local MW ($\omega _{L}$) fields were the input fields. First, the $\omega _{p}$ and $\omega _{c}$ fields pumped atoms from ground state $\left\vert 1\right\rangle $ to Rydberg state $\left\vert 3\right\rangle $. Then, as shown in Fig.~\ref{expFig}(b), the $\omega _{L}$ field induced atoms emitted in state $\left\vert 4\right\rangle $, and the signal MW field ($\omega _{S}$) excited atoms to $\left\vert 3\right\rangle$ and finally released a $\omega _{p}^{+}$ photon. This six-wave mixing process satisfied the condition $\omega _{p}^{+}=\omega _{p}+\omega _{c}-\omega_{c}+\omega _{L}-\omega _{S}$. Thus, $\omega _{p}^{+}=\omega _{p}+\delta_{s} $ with $\delta _{s}=\omega _{L}-\omega _{S}$, which is a positive sideband photon. Alternatively, as shown in Fig.~\ref{expFig}(c), the input signal MW field ($\omega _{S}$) induced atoms emitted at state $\left\vert 4\right\rangle $, and the strong $\omega _{L}$ field excited atoms to state $\left\vert 3\right\rangle $ and finally released a $\omega _{p}^{-}$ photon, which satisfied the condition $\omega_{p}^{-}=\omega _{p}+\omega _{c}-\omega _{c}-(\omega _{L}-\omega _{S})$. $\omega _{p}^{-}=\omega _{p}-\delta _{s}$ is a negative sideband photon. In these two processes, both intensities of generated $\omega _{p}^{+}$ and $\omega _{p}^{-}$ fields depended on the $\omega _{S}$ signal MW field; however, the $\omega _{p}^{-}$ field was stronger because the signal MW photon played different roles in these two processes. The signal MW photon acted as the seed of the stimulated emission process for the negative sideband generation and as linear absorption field for the positive sideband generation. Moreover, the wave vector mismatch $\Delta \bm{k}_\pm = \bm{k}_\pm -\bm{k}_p \mp \left(\bm{k}_L - \bm{k}_S\right)$, where $\bm{k}_\pm$ and $\bm{k}_Z$($Z \in \left\{p, L, S\right\} $) are the wave vectors of the corresponding field shown in Fig.~\ref{expFig}(a), had almost no effect on the efficiency of the six-wave-mixing process because $\abs{\Delta k_\pm l} \ll 1$ ($l= 5\,\mathrm{cm}$ was the length of the vapor cell) in our experiment geometry, which ensured wide input angle of the signal	MW detection in principle.

	In our system, the two generated sideband waves were mixed with a carrier wave (probe laser), forming a heterodyne detection. The frequency of the beat signal was $\delta _{s}$, which was the frequency difference between the two sideband waves and the carrier wave. The amplitude of the oscillating beat signal corresponded to the amplitude of the two generated sidebands, which directly reflected the strength of the electric field of the signal MW. The enhanced instantaneous bandwidth in our system was achieved by amplifying the negative sideband wave in the six-wave mixing process.

		To analyze dependence of the amplitude of beating signal on			parameters of the frequency-mixing processes and reveal the physical	mechanism of bandwidth enhancement, we began with a brief theoretical description of the atomic system response. In theory, for a weak probe and signal MW field, the beat signal amplitude $S(\delta _{s})$ oscillating at $\delta _{s}$ can be estimated as
		\begin{equation}
			S(\delta _{s})\propto \left\vert \left\langle \rho _{12}^{1}\right\rangle
			-\left\langle {\rho _{12}^{-1}}\right\rangle ^{\ast }\right\vert ,
			\label{feq}
		\end{equation}
		where the angular brackets denote the Doppler average, $\rho _{12}^{1}$ and $\rho _{12}^{-1}$ are the harmonic atomic coherence components of $\delta_s$ (i.e., $\rho=\sum_{n}\rho^n e^{in\delta_s t}$), which are described as
		\begin{eqnarray}
	     \rho_{12}^1 =&& \frac{i}{2} \frac{\Omega_p \abs{\Omega_c}^2\Omega_L\Omega_S^* }{G(\gamma_{12},\gamma_{13},\gamma_{14}) }  \nonumber \\
	    && \times\frac{\gamma_{14}}{G(\gamma_{12}+i\delta_s,\gamma_{13}+i\delta_s,\gamma_{14}+i\delta_s)}, \label{eq_6P} 	\\
	    \rho_{12}^{-1} =&& \frac{i}{2} \frac{\Omega_p \abs{\Omega_c}^2\Omega_L^*\Omega_S }{G(\gamma_{12},\gamma_{13},\gamma_{14})}   \nonumber \\
	    &&\times \frac{\gamma_{14}-i\delta_s}{G(\gamma_{12}-i\delta_s,\gamma_{13}-i\delta_s,\gamma_{14}-i\delta_s)},  \label{eq_6N}
        \end{eqnarray}	
		where $\gamma_{12} = \gamma+\Gamma_2/2+i\Delta_p$, $\gamma_{13} =\gamma+\Gamma_r/2+i(\Delta_p+\Delta_c)$, $\gamma_{14} =\gamma+\Gamma_r/2+i(\Delta_p+\Delta_c-\Delta_L)$, $\Gamma_2$, and $\Gamma_r$ are the spontaneous decay rates of the excited and Rydberg states, and $\gamma$ is the additional dephasing rate of atomic coherence. It should be noted that we had $\Gamma_2=2\pi\times6.07\,\mathrm{MHz}$, $\Gamma_r=2\pi\times2.4\,\mathrm{kHz}$, and the maximum value $\gamma=2.76\,\mathrm{MHz}$ in our system. The details of the coefficients $G(\gamma_{12},\gamma_{13}, \gamma_{14})$ and the theoretical model are presented in \cite{sup}, including Refs.~\cite{theory_arc, theory_Eq, theory_floquet, decay_atom, decay_Rydberg}. Eq.~(\ref{feq}) was used to calculate the theoretical response curves. According to Eqs.~(\ref{feq}--\ref{eq_6N}), it is clear that the amplitude of the beat signal $S(\delta_s)$ of the two frequency-mixing processes depends on the Rabi frequencies of the coupling laser and local MW.
		
	\begin{figure}
		\includegraphics[width=8.6cm]{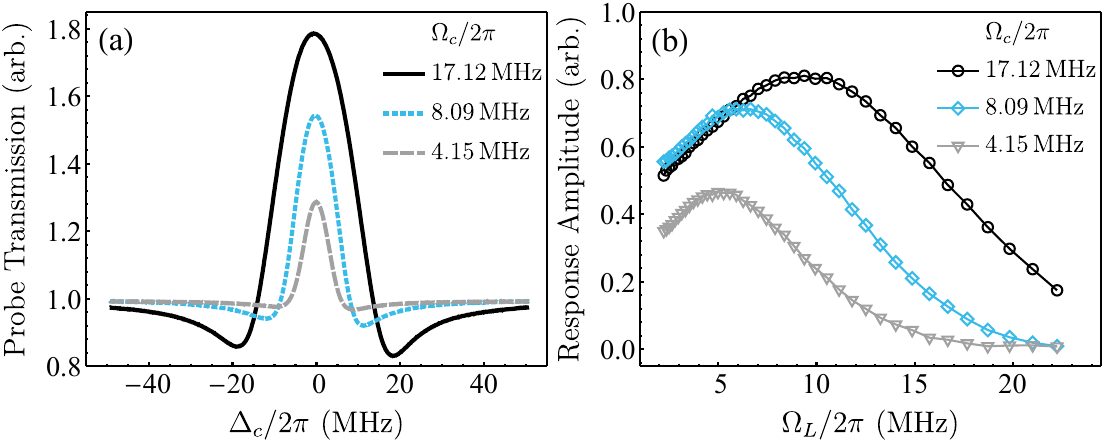}
		\caption{ \label{optFig} Optimization of Rydberg MW sensor. (a) Probe transmission signals (EIT) as a function of coupling detuning without MWs for different coupling Rabi frequencies. (b) Response amplitude of the Rydberg sensor at $\delta_s = 100\,\mathrm{kHz}$ as a function of Rabi frequency of local MW field for different coupling Rabi frequencies. The generator output power of signal MW was $-50\,\mathrm{dBm}$ for all conditions. The response amplitude, corresponding to the beat signal oscillated at $\delta_s$, was measured using a spectrum analyzer. }
	\end{figure}

	In the experiment, we first optimized the local MW power for each coupling Rabi frequency to obtain the peak response of the system for the	signal MW. Fig.~\ref{optFig}(a) shows the measured EIT spectrum at different coupling Rabi frequencies, and Fig.~\ref{optFig}(b) shows the corresponding amplitude of the beat signal (i.e., the response amplitude for a small signal MW) versus the local MW Rabi frequency for $\delta_{s}=100\,\mathrm{kHz}$. It can be clearly observed that there was an optimized local MW power in which the response amplitude was maximum at a certain coupling Rabi frequency, and the peak response amplitude increased with the coupling Rabi frequency $\Omega _{c}$, indicating that the increase in the EIT signal exceeded the decrease caused by power broadening when $\Omega _{c}$ was smaller than $2\pi \times 17.12\,\mathrm{MHz}$. We measured the sensitivity and bandwidth of the weak MW signal under peak response conditions.

\begin{figure}
	\includegraphics[width=8.6cm]{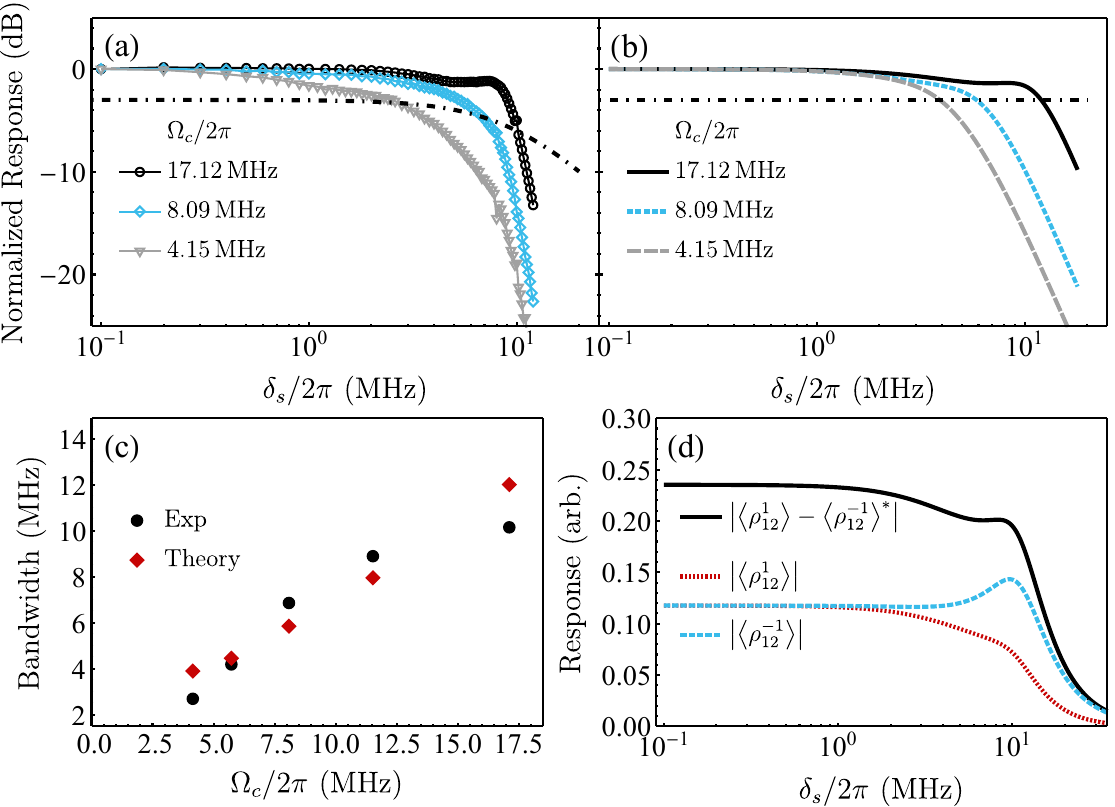}
	\caption{\label{bandFig} Instantaneous bandwidth of Rydberg MW sensor. (a) Experimental normalized response as a function of $\delta_s$ for different coupling Rabi frequencies. The measured response signal was normalized to the signal measured at $\delta_s = 100\,\mathrm{kHz}$. All response signals were measured in optimum conditions as those in Fig.~\ref{optFig}(b). The black dot-dashed line denotes the response of $-3\,\mathrm{dB}$ plus the attenuation of a first-order low-pass filter of $10\,\mathrm{MHz}$. (b) Theoretical normalized frequency response curve using Eq.~(\ref{feq}). The dot-dashed line denotes the $-3\,\mathrm{dB}$ line. (c) Comparison between experimental and theoretical instantaneous bandwidths for the different coupling Rabi frequencies. (d) Theoretical transient response contributed by two sideband waves with $\Omega_c =2\pi\times 17.12\,\mathrm{MHz}$. The contributions of the two sideband waves were estimated using $\abs{\avg{\rho_{12}^1}}$ (red dotted curve) for $\omega_p^+$ and $\abs{\avg{\rho_{12}^{-1}}}$ (cyan dashed curve) for $\omega_p^-$. The black solid curve denotes the total contribution. }
\end{figure}
	
	Here we demonstrate the instantaneous bandwidth of the proposed system. The atomic transient response curves were measured by scanning the frequency difference $\delta_{s} $ of the two MWs from $0.1\,\mathrm{MHz}$ to $12\,\mathrm{MHz}$. Fig.~\ref{bandFig}(a) shows the experimentally normalized atomic transient response curves for different coupling Rabi frequencies, and Fig.~\ref{bandFig}(b) shows the corresponding theoretical curves. In Fig.~\ref{bandFig}(a), the probe transmission signals were recorded using an APD with a bandwidth of $10\,\mathrm{MHz}$; hence, the measured frequency range was limited to $12\,\mathrm{MHz}$. In addition, the actual $-3\,\mathrm{dB}$ instantaneous bandwidth of the system was estimated from the intersection point between the measured frequency response curve and $-3\,\mathrm{dB}$ line, considering the bandwidth of the detector. The attenuation caused by the APD response was approximated using a first-order low-pass filter with a bandwidth of $10\,\mathrm{MHz}$. In Fig.~\ref{bandFig}(a), we observe a tendency toward an increased instantaneous bandwidth as the coupling Rabi frequency increases. Specifically, we observed a small gain peak in the atomic transient response curve at $\Omega _{c}=2\pi \times 17.12\,\mathrm{MHz}$. In Fig.~\ref{bandFig}(c), we plotted the experimental results versus the coupling Rabi frequency. The experimental instantaneous bandwidth of the system was $2.7\,\mathrm{MHz}$ for $\Omega _{c}=2\pi \times 4.15\,\mathrm{MHz}$, and increased to $10.2\,\mathrm{MHz}$ for $\Omega _{c}=2\pi \times 17.12\,\mathrm{MHz}$, which was significantly larger than the limit $\Gamma_{2}/2+\gamma \approx 5.8\,\mathrm{MHz}$ (the total ground-state dephasing rate) predicted in Ref.~\cite{b_band4}. The gain peak is crucial for enhancing the instantaneous bandwidth to overcome this limitation. The corresponding theoretical data are shown in Figs.~\ref{bandFig}(b) and (c). The theoretical predictions agree with the experimental data well, confirming the validity of the following analysis of the physical mechanism of bandwidth enhancement.

	To further investigate the origin of the enhanced instantaneous bandwidth and gain peak, we studied the contributions of the two generated sideband waves to the amplitude response, as shown in Fig.~\ref{bandFig}(d). In the low-frequency regime (i.e., $\delta _{s}/2\pi <1\,\mathrm{MHz}$), the contributions of the two generated sideband waves to the Rydberg receiver were almost identical. In addition, Fig.~\ref{bandFig}(d) clearly shows that in the high-frequency regime (i.e., $\delta _{s}/2\pi >1\,\mathrm{MHz}$), the contribution of the positive sideband wave ($\omega _{p}^{+}$) started to decrease, whereas that of the negative sideband wave ($\omega _{p}^{-}$) exhibited a large gain peak induced by a strong coupling laser, resulting in an overall enhanced instantaneous bandwidth. The distinct behaviors of the two sideband waves can be attributed to the roles of the signal MW field in the two six-wave mixing processes, namely, the seed field for stimulated emission and population pumping.
	
	\begin{figure}
		\includegraphics[width=8.6cm]{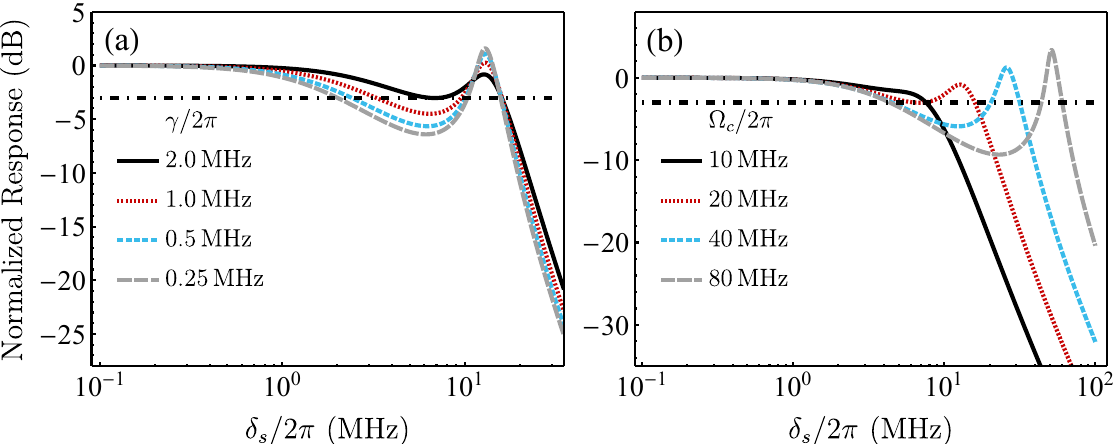}
		\caption{\label{disFig}Theoretical analysis of instantaneous bandwidth. (a) Role of the relaxation rate $\gamma$ on frequency response with coupling Rabi frequency $\Omega_c = 2\pi\times20\,\mathrm{MHz}$. For each relaxation rate $\gamma$, $\Omega_L$ was optimized to achieve peak response of signal MW. (b) Normalized frequency response for strong coupling laser with $\gamma/2\pi = 2.0\,\mathrm{MHz}$. The black dot-dashed lines in (a) and (b) denote $-3\,\mathrm{dB}$ response.}
	\end{figure}

	Furthermore, based on our theoretical model, we analyzed the bandwidth range that the system could reach. In theory, the atomic transient response highly depends on the dephasing rate and the coupling Rabi frequencies. In Fig.~\ref{disFig}(a), the theoretical frequency response curves at different dephasing rates clearly show the existence of a high-frequency response peak with an amplitude greater than $0\,\mathrm{dB}$ and a low-frequency response valley with an amplitude less than $-3\,\mathrm{dB}$, particularly at a low $\gamma$ rate. Fig.~\ref{disFig}(b) shows the theoretical frequency responses at different coupling Rabi frequencies $\Omega_{c}$. A response peak and valley appeared, and the frequency position of the peak increased with the coupling Rabi frequency. These two figures show that the dephasing rate $\gamma$ played a dominant role in the low-frequency response, whereas the strong coupling laser produced a high-frequency response peak.  A larger relaxation rate resulted in a flatter and broader bandwidth. In addition, it should also be noted that flat instantaneous bandwidths of up to several hundred megahertz could be easily obtained at the expense of sensitivity owing to the Doppler broadening effect in the copropagating experimental geometry of laser beams, as it effectively increased the relaxation rate $\gamma$. This study focused on simultaneously achieving both high sensitivity and broad instantaneous bandwidth. Under the premise of ensuring high sensitivity, our system achieved an experimental flat frequency range of up to 10.2 MHz, and theoretical analysis suggested that a high-frequency response of up to 50 MHz could be achieved using a large coupled Rabi frequency.
	
	\begin{figure}
		\includegraphics[width=7cm]{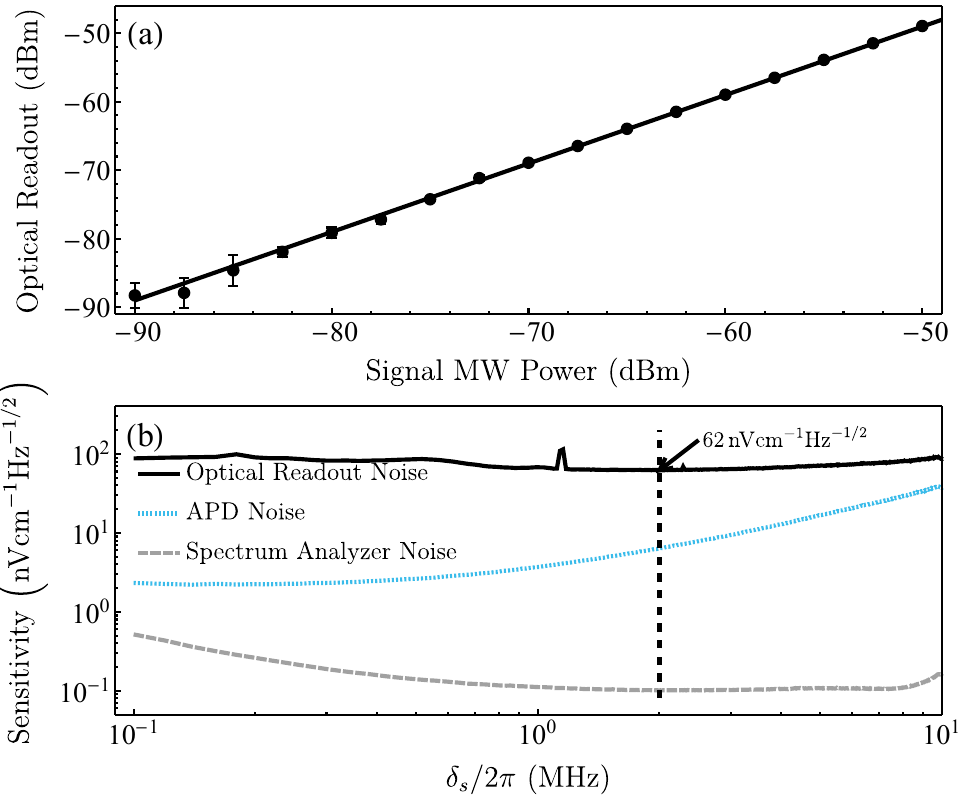}
		\caption{\label{senFig} Sensitivity of Rydberg MW sensor. (a) Optical readout of Rydberg superhet using spectrum analyzer as a function of the applied power of the signal MW from analog signal generator. The spectrum analyzer was in zero-span mode with $1$-Hz resolution bandwidth. The error bar denotes the standard deviation of measurement in $10$-s record. The solid line denotes the linear fit of the experimental data. (b) Noise limited sensitivity. The optical readout noise (black line) is the total noise measured without signal MW. The noise floor of APD and spectrum analyzer are shown in cyan dotted and gray dashed curves, respectively. The vertical black dashed line denotes $\delta_s/2\pi = 2\,\mathrm{MHz}$.}
	\end{figure}

	Next, we evaluated the sensitivity of our system by considering the noise level and signal response. Fig.~\ref{senFig}(a) shows the measured amplitude of the beat signal as a function of the applied signal MW power. The sensor performance was measured at optimal $\Omega _{c}=2\pi \,\times 17.12\,\mathrm{MHz}$, as shown in Fig.~\ref{optFig}(b). The optical readout signal exhibited excellent linearity with the signal MW power. Subsequently, the electric field amplitude of the signal MW field was deduced from the experimental data of the calibrated MW Rabi frequencies \cite{sup}. Thus, we deduced the sensitivity of our system from the noise measurement, where signal-to-noise ratio equal to one. Fig.~\ref{senFig}(b) shows the noise-limited sensitivity of the Rydberg superhet. It should be noted that we considered the frequency-response effect shown in Fig.~\ref{bandFig}(a) to accurately determine the sensitivity of this system. The best sensitivity of our setup was estimated to be $62\,\sUnit$ at $\delta _{s}/2\pi =2\,\mathrm{MHz}$. The APD noise floor was lower than the optical noise floor, indicating that the current sensitivity was mainly limited by the system noise. Higher sensitivity may be achieved through homodyne detection to eliminate common-mode laser noise.
	
	In summary, we demonstrated high-performance Rydberg MW electrometry, which has practical advantages: a high sensitivity of $62\,\sUnit$ and a broad instantaneous bandwidth of up to $10.2\, \mathrm{MHz}$. The observed frequency response, which exhibited an enhanced high-frequency gain in the strong-coupling case, was attributed to the amplification of one generated sideband wave in the six-wave mixing process of the Rydberg superheterodyne receiver. Our theoretical analysis showed that the detection of a superheterodyne signal at frequencies of up to $50\,\mathrm{MHz}$ can also be achieved with a large coupling Rabi frequency. For the first time, our results revealed an enhanced instantaneous response and realized broadband and high-sensitivity microwave measurements, promoting future practical applications.

\begin{acknowledgments}
This study was financially supported by the National Natural Science Foundation of China (Grants No. 12174409 and No. 61835013). 

\end{acknowledgments}


\bibliography{reference}

\end{document}


	
	\title{Supplemental Materials: High-Sensitive Microwave Electrometry with Enhanced Instantaneous Bandwidth}
	
	
	\author{Bowen Yang}
	\affiliation{Key Laboratory of Quantum Optics and Center of Cold Atom Physics, Shanghai Institute of Optics and Fine Mechanics, Chinese Academy of Sciences, Shanghai 201800, China}
	\affiliation{Center of Materials Science and Optoelectronics Engineering, University of Chinese Academy of Sciences, Beijing 100049, China}

	\author{Yuhan Yan}
	\affiliation{Key Laboratory of Quantum Optics and Center of Cold Atom Physics, Shanghai Institute of Optics and Fine Mechanics, Chinese Academy of Sciences, Shanghai 201800, China}
	\affiliation{Department of Physics and Electronic Science, East China Normal University, Shanghai 200062, China}
	
	\author{Xuejie Li}
	\affiliation{Key Laboratory of Quantum Optics and Center of Cold Atom Physics, Shanghai Institute of Optics and Fine Mechanics, Chinese Academy of Sciences, Shanghai 201800, China}
	\affiliation{Department of Physics and Electronic Science, East China Normal University, Shanghai 200062, China}
	
	\author{Ling Xiao}
	\affiliation{Key Laboratory of Quantum Optics and Center of Cold Atom Physics, Shanghai Institute of Optics and Fine Mechanics, Chinese Academy of Sciences, Shanghai 201800, China}

	\author{Xiaolin Li}
	\affiliation{School of Physics, East China University of Science and Technology}
	
	\author{L. Q. Chen}
	\email[]{lqchen@phy.ecnu.edu.cn}
	\affiliation{Department of Physics and Electronic Science, East China Normal University, Shanghai 200062, China}
	\affiliation{Shanghai Branch, Hefei National Laboratory, Shanghai 201315, China}
	
	\author{Jianliao Deng}
	\email[]{jldeng@siom.ac.cn}
	\affiliation{Key Laboratory of Quantum Optics and Center of Cold Atom Physics, Shanghai Institute of Optics and Fine Mechanics, Chinese Academy of Sciences, Shanghai 201800, China}
	
	\author{Huadong Cheng}
	\email[]{chenghd@siom.ac.cn}
	\affiliation{Key Laboratory of Quantum Optics and Center of Cold Atom Physics, Shanghai Institute of Optics and Fine Mechanics, Chinese Academy of Sciences, Shanghai 201800, China}
	\affiliation{Center of Materials Science and Optoelectronics Engineering, University of Chinese Academy of Sciences, Beijing 100049, China}
	

	
	\date{\today}

	
	\maketitle


\section{Theoretical Model}
	
	The time-evolution equation of the density matrix element of the atomic system driven by lasers and MW sources in the rotating frame is obtained from the master equation \cite{theory_Eq}
	\begin{equation}
		\dot{\rho} = -i\left[H_0,\rho\right] -i\left[H_S(t),\rho\right] + \mathcal{L}_\rho,
		\label{eq1}
	\end{equation}
	where $H_0$ is the Hamiltonian of the atomic system without a signal MW, $H_S$ is the perturbation Hamiltonian driven by a small signal MW, and $\mathcal{L}_\rho$ is the Lindblad operator that accounts for the relaxation terms. The Hamiltonians $H_0$ and $H_s$ for the four-level system shown in Fig. 1(b) are described as follows:
	\begin{eqnarray}
		H_0 =&& -\Delta_p \mstate{2} -\left(\Delta_p+\Delta_c\right) \mstate{3} -\left(\Delta_p+\Delta_c - \Delta_L \right) \mstate{4} \nonumber	\\
		&&+\frac{1}{2}\left(\Omega_p\Mstate{1}{2} + \Omega_c \Mstate{2}{3}  + \Omega_L^* \Mstate{3}{4} +\mathrm{H.c.}\right),	\\
		H_S =&& \frac{1}{2}\left( \Omega_S^* e^{i\delta_s t}\Mstate{3}{4} + \mathrm{H.c.} \right),
	\end{eqnarray}
	where $\Mstate{i}{j} = \left|i \right\rangle \left\langle j\right| $ is the projection operator; $\Delta_p = \omega_p - \omega_{1,2} $, $\Delta_c = \omega_c - \omega_{2,3}$, and $\Delta_L = \omega_L - \omega_{3,4}$ are the frequency detunings of the probe, coupling, and local MW fields, respectively; $\Omega_Z$ ($Z \in\left\{ p, c, L, S\right\}$) is the Rabi frequency of corresponding field; and $\delta_s = \omega_L -\omega_S$ is the frequency difference of two MWs. Specifically, for a weak probe-field approximation, the evolution of the off-diagonal density matrix elements can be simplified using Eq.~(\ref{eq1}) in the form
	\begin{equation}
		\dot{\bm{X}} = \left(A+ \Omega_S^* B_1 e^{i\delta_s t} + \Omega_S B_{-1}e^{-i\delta_s t} \right)\bm{X} + \bm{C}, \label{eq_dif}
	\end{equation}
	where $\bm{X}$, $\bm{C}$, $A$, $B_1$ and $B_{-1}$ are expressed as
	\begin{eqnarray}
		\bm{X} =&& \left(\rho_{12}, \rho_{13}, \rho_{14}\right)^T \text{, } \bm{C} = \left(i\Omega_p/2, 0, 0\right)^T,	\\
		A =&& \begin{pmatrix}
			-\gamma_{12}	& i\Omega_c^*/2	& 0	\\
			i\Omega_c/2	& -\gamma_{13}	& i\Omega_L/2	\\
			0			& i\Omega_L^* /2	&-\gamma_{14}	
		\end{pmatrix},	\\
		B_1 =&& \begin{pmatrix}
			0		& 0		&0	\\
			0		& 0		& 0 \\
			0		&i/2		&0		
		\end{pmatrix}
		\text{ and }
		B_{-1} = \begin{pmatrix}
			0		& 0		&0	\\
			0		& 0		&i/2 \\
			0		&0		&0		
		\end{pmatrix},
	\end{eqnarray}
	where $\gamma_{12} = \gamma+\Gamma_2/2+i\Delta_p$, $\gamma_{13} = \gamma+\Gamma_r/2+i(\Delta_p+\Delta_c)$, and $\gamma_{14} = \gamma+\Gamma_r/2+i(\Delta_p+\Delta_c-\Delta_L)$. Note that $\Gamma_2$ and $\Gamma_r$ are the decay rates of the excited and Rydberg states, chosen as $\Gamma_2 = 2\pi\times6.07\,\mathrm{MHz}$ and $\Gamma_r = 2\pi\times2.4\,\mathrm{kHz}$, respectively, based on ~\cite{decay_atom} and~\cite{decay_Rydberg}. This simple model is consistent with that in the Supplementary Material of \cite{theory_t_model}. In our simple model, the relaxation rate $\gamma$ accounts for the finite interaction time between the laser beams and atoms owing to the atomic thermal motion and collision between atoms and other non-transit sources, and is estimated using the theoretical fit (see Section.~\ref{SexpData}). In our experiment, $\gamma/2\pi$ was estimated as $1.31-2.76\,\mathrm{MHz}$.
	
	The desired solution of Eq.~(\ref{eq_dif}) is periodic. Thus, we performed a Floquet analysis of the system by expanding $\bm{X}(t)$ into the harmonics of frequency $\delta_s$ \cite{theory_floquet}. Then, we used the following equation:
	\begin{equation}
		\bm{X}(t) = \sum_{n=-\infty}^{\infty} \bm{X}_n e^{i n \delta_s t}, \label{eq_expand}
	\end{equation}
	where $\bm{X}_n$ is expressed as follows:
	\begin{equation}
		\bm{X}_n = \left(\rho_{12}^n, \rho_{13}^n, \rho_{14}^n \right)^T.
	\end{equation}
	Substituting Eq.(\ref{eq_expand}) into Eq.~(\ref{eq_dif}) and considering the stationary case ($\dot{\bm{X}}_n = 0$), we obtained the recursion equation of $\bm{X}_n$, which is expressed as
	\begin{eqnarray}
		\left(A-in\delta_s I\right)\bm{X}_n + \Omega_s^* B_1 \bm{X}_{n-1} 
		+ \Omega_s B_{-1} \bm{X}_{n+1} 
		+ \bm{C}\delta_{n,0} = 0,
	\end{eqnarray}
	where $I$ denotes an identity matrix. An approximate solution was obtained by eliminating the high-harmonic terms of $\abs{n}\ge 2$ because of the small signal MW field. The first-order approximation solution yields
	\begin{widetext} 
		\begin{eqnarray}
			\rho_{12}^0 =&& \frac{i\Omega_p}{2} \frac{ 4\gamma_{13}\gamma_{14}+\abs{\Omega_L}^2}{G(\gamma_{12},\gamma_{13},\gamma_{14})},	\label{eq_p0}\\
			\rho_{12}^1 =&& \frac{i}{2} \frac{\gamma_{14}\Omega_p \abs{\Omega_c}^2\Omega_L\Omega_S^* }{G(\gamma_{12},\gamma_{13},\gamma_{14}) \label{eq_0} G(\gamma_{12}+i\delta_s,\gamma_{13}+i\delta_s,\gamma_{14}+i\delta_s)}, \label{eq_6P} 	\\
			\rho_{12}^{-1} =&& \frac{i}{2} \frac{(\gamma_{14}-i\delta_s)\Omega_p \abs{\Omega_c}^2\Omega_L^*\Omega_S }{G(\gamma_{12},\gamma_{13},\gamma_{14}) G(\gamma_{12}-i\delta_s,\gamma_{13}-i\delta_s,\gamma_{14}-i\delta_s)},  \label{eq_6N}
		\end{eqnarray}
	\end{widetext}
	where the coefficient $G(\gamma_{12},\gamma_{13},\gamma_{14})$ is defined as
	\begin{eqnarray}
		&&G(\gamma_{12},\gamma_{13},\gamma_{14}) \nonumber \\
		\quad&&= \gamma_{14} \abs{\Omega_c}^2 +\gamma_{12}\left(4\gamma_{13}\gamma_{14} + \abs{\Omega_L}^2\right). \label{eq_G}
	\end{eqnarray}
	Eqs.(\ref{eq_6P}) and (\ref{eq_6N}) describe the six-wave mixing processes of the two generated sidebands of the carrier–probe field.	Moreover, the density matrix element $\rho_{12}$ must be Doppler-averaged. Thus, we obtained
	\begin{equation}
		\avg{\rho_{12}^{0,\pm 1} } = \frac{1}{\sqrt{\pi} v_p}\int_{-\infty}^{\infty} e^{-v^2/v_p^2}\rho_{12}^{0,\pm1}(\Delta_p^\prime,\Delta_c^\prime) dv, \label{eq_Doppler}
	\end{equation}
	where $v_p = \sqrt{2k_B T/m}$ denotes the most probable velocity of the atoms. In this integration, the probe and coupling laser detunings were modified as $\Delta_p^\prime = \Delta_p -k_p v$ and $\Delta_c^\prime = \Delta_c + k_c v$, respectively, where $k_p$ and $k_c$ are the wave vectors of the probe and coupling lights, respectively.
	
	The measured superheterodyne signal was related to the variation in the probe transmission. In this case, the observed signal $S$ was proportional to the imaginary part of the atomic coherence $\rho_{12}$ or the beat signal between the carrier probe field and the two generated sideband waves. Thus, we have
	\begin{equation}
		S \propto \mathrm{Im}\left(\avg{\rho_{12}^1} e^{i\delta_s t} + \avg{\rho_{12}^{-1}} e^{-i\delta_s t}\right).
	\end{equation} 
	Thus, the beat signal amplitude $S(\delta_s)$ that oscillates at $\delta_s$ can be expressed as follows:
	\begin{equation}
		S(\delta_s) \propto \abs{\avg{\rho_{12}^1} - \avg{{\rho_{12}^{-1}}}^*}.
		\label{feq}
	\end{equation}
	Eq.~(\ref{feq}) was used to calculate the theoretical response curves.
	
\section{Experimental Parameters \label{SexpData}}

	The relevant electric dipole moments are listed in Table\ref{rabi}. The transition dipole moments for the coupling and MW fields were estimated using the alkali Rydberg calculator package \cite{theory_arc}. 
	\begin{table}[htbp]
		\caption{\label{rabi}Electric transition dipole moments}
		\begin{ruledtabular}
			\begin{tabular}{llcc}
				Field	& Transition\footnote{All fields are $\pi$ polarized}	& Waist\footnote{The waist is the geometric average value of waists in $X$-direction and $Y$-direction.} ($\mathrm{\mu m}$) 		& Dipole moment ($ea_0$)\footnote{Here $e$ is the elementary charge and $a_0$ is the Bohr Radius.} 	\\
				Probe	 	& $\state{5S_{1/2}}\to \state{5P_{3/2}}$	& 78.66	&2.44			\\
				Coupling 	& $\state{5P_{3/2}}\to \state{51D_{5/2}}$	& 100.24	&0.012			\\
				signal MW	&$\state{51D_{5/2}}\to \state{52P_{3/2}}$ 	& NA	&1640.184		\\
				local MW	&$\state{51D_{5/2}}\to \state{52P_{3/2}}$ 	& NA	&1640.184		\\
			\end{tabular}
		\end{ruledtabular}
	\end{table}

	The Rabi frequencies of laser beams are typically determined using the following equation:
	\begin{equation}
		\Omega = \sqrt{\frac{4P}{\epsilon_0 c\pi w^2}}\frac{d_e}{\hbar},
		\label{calRabi}
	\end{equation}
	where $P$ is the laser power, $w$ is the Gaussian beam $1/e^2$ radius, $\epsilon_0$ is the permittivity of vacuum, $c$ is the speed of light, $d_e$ is the dipole moment of the corresponding transition, and $\hbar$ is Planck’s constant. Because of the Gaussian intensity distribution of the laser beam, the effective Rabi frequency perceived by the atoms in the region of action differed from the peak Rabi frequency. We derived the effective coupling Rabi frequency and relaxation rate $\gamma$ by theoretically fitting the experimental EIT spectrum.
	
	According to Eqs.~(\ref{eq_0}--\ref{eq_Doppler}) and setting $\Omega_L=0$, the measured EIT probe transmission is as follows:  
	\begin{equation}
		\ln \frac{P_{\text{out}}}{P_{\text{in}}} = \frac{\beta}{\sqrt{\pi}v_p}\int_{-\infty}^{\infty}e^{v^2/v_p^2} \mathrm{Im}\left(\frac{\rho_{12}(\Delta_p^\prime,\Delta_c^\prime)}{\Omega_p}\right),
		\label{eq_eit}
	\end{equation}
	where $P_{\text{out}}$ and $P_{\text{in}}$ are the optical powers of the transmitted and incident probe lights, respectively, and $\beta$ is the scale	factor related to the atomic density and physical constants, which was estimated using the following relation:
	\begin{equation}
		OD = \frac{\beta}{\sqrt{\pi}v_p}\int_{-\infty}^{\infty}e^{v^2/v_p^2} \frac{1}{2} \frac{1}{{\Delta_p^\prime}^2 +\left(\Gamma_2/2 \right)^2},
	\end{equation}
	where $OD$ denotes the optical depth of the atomic medium. In our study, $OD\approx1.16$ corresponded to $\beta \approx 406.8$.
	
\begin{figure*}
		\includegraphics[width=17.2cm]{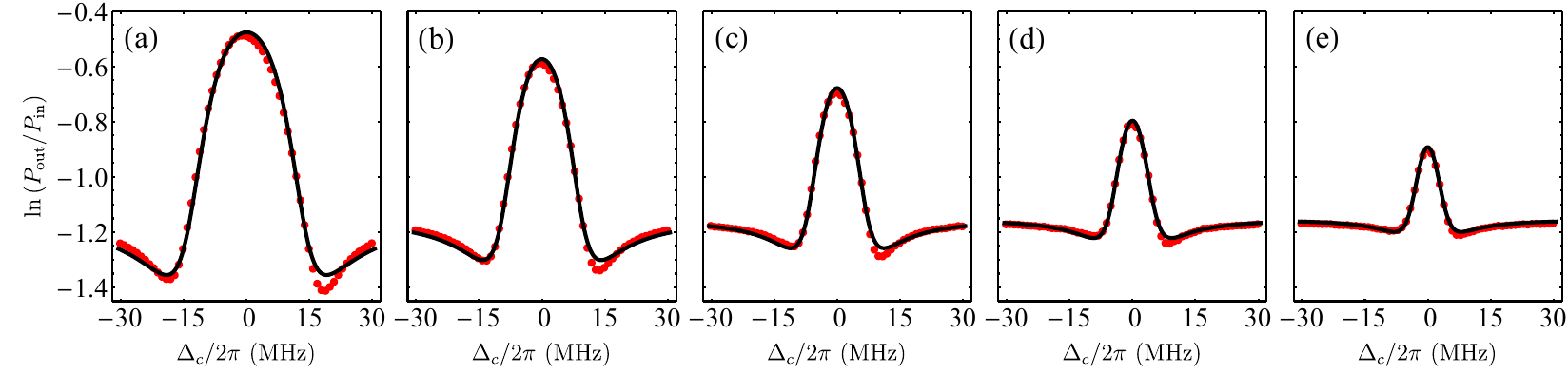}
		\caption{ \label{figS0} EIT signals with different coupling laser power levels. (a), (b), (c), (d) and (e) correspond to the coupling laser power levels of $536\,\mathrm{mW}$, $250\,\mathrm{mW}$, $124.7\,\mathrm{mW}$, $62.4\,\mathrm{mW}$, and $31.2\,\mathrm{mW}$, respectively. The red dots indicate the experimental data obtained at probe power $125.4\,\mathrm{nW}$. Experimental results (red dots) were fitted to Eq.~(\ref{eq_eit}) (black solid curves).}
	\end{figure*}

	Fig.~\ref{figS0} shows the measured spectrum and the corresponding theoretical fit using Eq.~(\ref{eq_eit}). Table.\ref{fit} presents the coupling Rabi frequencies $\Omega_c$ and relaxation rate $\gamma$ obtained through theoretical fit. The results show that the Rabi frequencies obtained by theoretical fitting were smaller than the peak Rabi frequencies calculated using Eq.~(\ref{calRabi}). In this study, the fitting results were used to characterize the coupling Rabi frequency. It was found that the relaxation rate $\gamma$ increased with an increase in the coupling power, which may be due to the enhanced collision rate and additional power broadening effect.
	
\begin{figure*}
		\includegraphics[width=17.2cm]{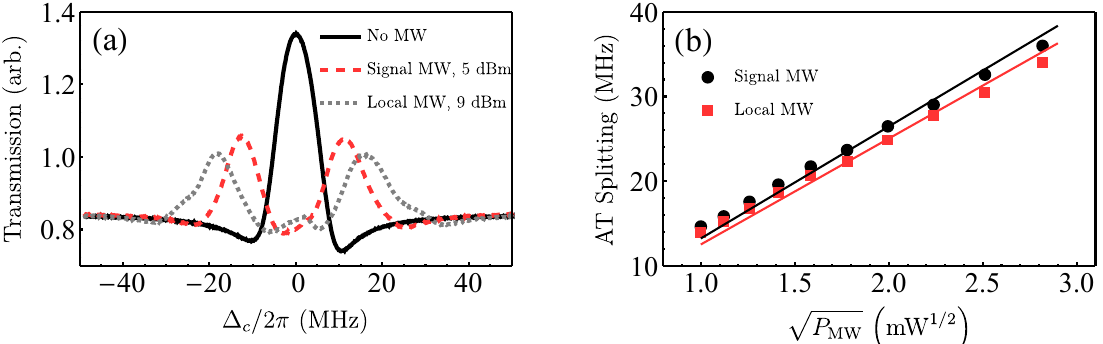}
		\caption{ \label{figS1} (a) Probe transmission signals as functions of coupling detuning with and without MWs. All data traces were averaged $40$ times and filtered by a digital low-pass filter. (b) Measured AT splitting data as a function of the square root of signal or local MW power. The solid curve denotes the linear fit of experiment data in the form $\Delta f_{\mathrm{AT}}=\alpha_{S,L}\sqrt{P_{\mathrm{MW}}}$.}
	\end{figure*}

	\begin{table}[htbp]
		\caption{\label{fit}Coupling Rabi Frequencies $\Omega_c$ and relaxation rates $\gamma$}
		\begin{ruledtabular}
			\begin{tabular}{lllll}
				Figure&	Coupling Power & $\gamma/2\pi$ (Fit results) & $\Omega_c/2\pi$ (Fit results) & $\Omega_c/2\pi$ (calculated by Eq.~(\ref{calRabi}))	\\
				(a) & $536\,\mathrm{mW}$ & $2.76(5)\,\mathrm{MHz}$ & $17.12(10)\,\mathrm{MHz}$ & $23.95\,\mathrm{MHz}$	\\
				(b)	& $250\,\mathrm{mW}$ & $2.13(3)\,\mathrm{MHz}$ & $11.53(7)\,\mathrm{MHz}$ &	$16.35\,\mathrm{MHz}$			\\
				(c) & $124.7\,\mathrm{mW}$ & $1.75(3)\,\mathrm{MHz}$ & $8.09(5)\,\mathrm{MHz}$ &	$11.55\,\mathrm{MHz}$					\\
				(d) & $62.4\,\mathrm{mW}$ & $1.52(3)\,\mathrm{MHz}$ & $5.73(4)\,\mathrm{MHz}$ &	$8.17\,\mathrm{MHz}$					\\
				(e) & $31.2\,\mathrm{mW}$ & $1.31(3)\,\mathrm{MHz}$ & $4.15(4)\,\mathrm{MHz}$ &	$5.78\,\mathrm{MHz}$					
			\end{tabular}
		\end{ruledtabular}
	\end{table}
		The MW Rabi frequency was calibrated using Autler-Townes (AT) splitting. Fig.~\ref{figS1} shows examples of the EIT-AT spectra and measured AT splitting of the signal and local MWs versus the applied power. The linear fits resulted in a free-space transmitted gain $\alpha_S = 13.22(16)\,\mathrm{MHz/mW^{1/2}}$ for the signal MW and $\alpha_L = 12.51(17)\,\mathrm{MHz/mW^{1/2}}$ for the local MW. Furthermore, the corresponding electric field amplitude for a given MW power was deduced using the transition dipole moment. The relationship between the electric field amplitude and MW power was $6.30(8)\,\mathrm{mV/cm/mW^{1/2}}$ for the signal MW and $5.96(8)\,\mathrm{mV/cm/mW^{1/2}}$ for the local MW.
	
\bibliography{reference}